    \documentstyle[epsf]{article} \catcode`\@=11
    \def\section{\@startsection{section}{1}{\z@}%
    {-3.5ex plus -1ex minus -.5ex}{1.5ex plus.3ex}{\bf }}
    \def\subsection{\@startsection{subsection}{1}{\z@}%
    {-3.5ex plus-1ex minus-.5ex}{1.5ex plus.3ex}{\bf }} 
    
\begin{document}
    \hfill\parbox{4.77cm}{\Large\centering Annalen\\der
    Physik\\[-.2\baselineskip] {\small \underline{\copyright\ Johann
    Ambrosius Barth 1998}}} \vspace{.75cm}\newline{\Large\bf
Energy level statistics at the metal-insulator transition in the
Anderson model of localization with anisotropic hopping
    }\vspace{.4cm}\newline{\bf   
F.\ Milde and R.A.\ R{\"o}mer
    }\vspace{.4cm}\newline\small
Institut f{\"u}r Physik, Technische Universit{\"a}t, D-09107
  Chemnitz, Germany
    \vspace{.2cm}\newline 
printed: \today
    \vspace{.4cm}\newline\begin{minipage}[h]{\textwidth}\baselineskip=10pt
    {\bf  Abstract.}
    Recently, a metal-insulator transition (MIT) was found in the
    an\-iso\-tropic Anderson model of localization by transfer-matrix
    methods (TMM) \cite{AEbenen1,AKetten,Zamb96}. This MIT has been
    also investigated by multifractal analysis (MFA) \cite{Multi} and
    the same critical disorders $W_c$ have been obtained within the
    accuracy of the data. We now employ energy level statistics (ELS)
    to further characterize the MIT. We find a crossover of the
    nearest-neighbor level spacing distribution $P(s)$ from GOE
    statistics at small disorder indicating metallic behavior to the
    Poisson distribution at large disorder characteristic for
    localized states.  An analysis of the system size dependence of
    the spectral rigidity $\Delta_3(L)$ confirms the values of $W_c$
    found in Ref.\ \cite{Zamb96,Multi}.
    \end{minipage}\vspace{.4cm} \newline {\bf  Keywords:}
metal-insulator transition, disordered systems, anisotropic systems
    \newline\vspace{.2cm} \normalsize
\section{Introduction}

Recent TMM studies \cite{AEbenen1,AKetten,Zamb96} of the anisotropic
Anderson model show that an MIT exists even for strong hopping
anisotropy $\gamma$.  The values of the critical disorder in the band
center were found to follow a power law $W_c\propto(1-\gamma)^\beta$
independent of the orientation of the quasi-1D bar. $\beta$ was argued
to be independent of the strength of the anisotropy. This is supported
by multifractal analysis of the eigenfunctions \cite{Multi}, where the
system size dependence of the singularity spectra is used to determine
$W_c$. In order to check this further, we employ energy level
statistics as another independent method. ELS is based on random
matrix theory \cite{Mehta} and was successfully used to investigate
the MIT in the isotropic case \cite{Zara97,Hof93,Hof94}. Directly at
the MIT, the ELS was argued to be given by a universal distribution
$P_c(s)$, which should correspond to a ``critical ensemble''
\cite{Zara97} distinct from the case of a Gaussian orthogonal ensemble
(GOE) and also from the Poissonian case. However, recent results
\cite{Schweitz} show that $P_c(s)$ depends on the boundary conditions
and the shape of the samples considered.

In this work, we show that the MIT in the anisotropic systems can be
characterized conveniently by ELS and we find that the critical
disorders $W_c$ are in good agreement with the results of Refs.\ 
\cite{Zamb96,Multi}. We further find that $P(s)$ at the MIT in the
anisotropic systems depends on the anisotropy $\gamma$. Thus an
estimation of $W_c$ by use of the ``critical ensemble'' is incorrect.

\section{The anisotropic Anderson model of localization}

The Anderson Hamiltonian is given as \cite{And58}
\begin{equation}
  \label{Hand}
  H = \sum_{i} \epsilon_{i} | i \rangle\langle i | + \sum_{i \ne j}
  t_{ij} | i \rangle\langle j | \quad .
\end{equation}
We use a simple cubic lattice of size $N^3$ with orthonormal states $|
i \rangle$ at site $i=(x,y,z)$. The potential site energies
$\epsilon_{i}$ are random numbers, uniformly distributed in the
interval $[-W/2,+W/2]$.  The transfer integrals $t_{ij}$ are
restricted to nearest neighbors. They depend only on the spatial
direction, thus $t_{ij}=t_x$, $t_y$ or $t_z$.  We study the two cases
of: (i) {\em weakly coupled planes} with $t_x=t_y=1$, $t_z=1-\gamma$
and (ii) {\em weakly coupled chains} with $t_x=t_y=1-\gamma$, $t_z=1$.
The anisotropy parameter $\gamma$ ranges from $\gamma=0$, the
isotropic case, to $\gamma=1$ where the planes/chains are completely
uncoupled.

We use the Lanczos algorithm \cite{CulW85} to compute the spectrum of
$H$. It is well suited for the diagonalization of our sparse matrices
\cite{challenge} and allows us to compute all eigenvalues of $H$ for
system sizes of $N=48$ on a parallel machine within 60 hours. We use
50\% of the eigenvalues around the band center $E=0$ and average over
up to 400 realizations of the random potential, such that at least $2
\times 10^5$ eigenvalues are used for each set of parameters
$\{W,\gamma,N\}$.  Due to the large computational effort, we restrict
the systematic investigations to sizes up to $N=30$. For comparison
with predictions of random matrix theory, we unfold the spectra by
fitting cubic splines \cite{Hof93} to the integrated density of
states. This sets the mean-level spacing to one.  We then characterize
the local spectral fluctuations by means of the nearest neighbor level
spacing distribution $P(s)$ and the $\Delta_3$ statistics.  The latter
measures the rigidity of the spectra \cite{Mehta}.

\section{Results}

{\em Extended} states in a metal contribute to charge transport even
at $T=0$. The overlap of the extended states results in level
repulsion and their spectral properties are characterized by the GOE.
On the other hand, {\em localized} states cannot contribute to charge
transfer at $T=0$, resulting in insulating behavior. The energy levels
are uncorrelated, consequently the probability that energy levels are
close together is very high and the ELS is given by the Poissonian
statistics. Thus a change from the GOE behavior to Poissonian may
indicate the existence of an MIT.

\begin{figure}[th]
  {\epsfysize=6cm\epsfbox{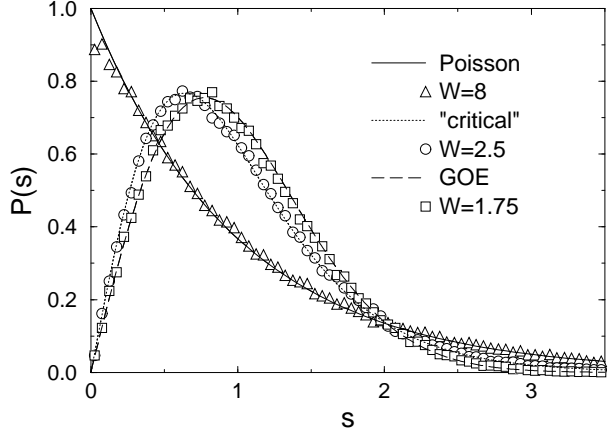}}
  \caption{\protect\small 
    $P(s)$ for weakly coupled chains with $N=24$ and $\gamma=0.9$.}
  \label{fig:p_s}
\end{figure}

As expected from the isotropic case, we find $P(s)$ to be close to the
GOE statistics at small disorder and close to Poisson statistics at
large disorder as shown in Fig.~\ref{fig:p_s} for an already quite
strong anisotropy.
For stronger anisotropy $\gamma$, the transition occurs at
smaller values of the disorder parameter $W$.

In an infinite system, there is a sharp transition from extended to
localized behavior at the MIT. However, in any finite system, the
characteristic lengths scales of the states close to the transition
will exceed the system size. Thus for a given $N$, one finds
characteristic deviations which result in a continuous change from GOE
to Poissonian statistics as $W$ is varied across the MIT.  Only
directly at $W_c$ the statistical properties are independent of $N$,
because of the scale invariance of the multifractal wave functions at
the MIT \cite{Schreib,Multi}.  In order to identify the extended,
critical and localized regimes and to determine the critical disorder
$W_c$ properly, we therefore examine the system size dependence of the
ELS.

\begin{figure}[th]
  {\epsfysize=6cm\epsfbox{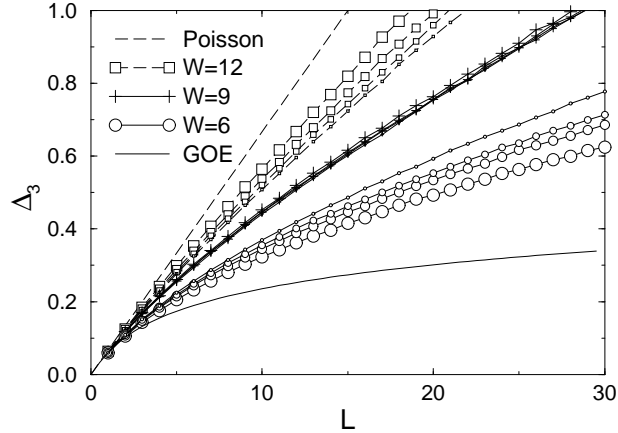}}
  \caption{\protect\small 
    $\Delta_3(L)$ for weakly coupled planes with $\gamma=0.9$ for
    $N=13,17,21,30$. Larger $N$ corresponds to larger symbol size.}
    \label{fig:d3}
\end{figure}

As an example we show in Fig.~\ref{fig:d3} the $\Delta_3$ statistics
for weakly coupled planes at $\gamma=0.9$ for $4$ system sizes $N$
ranging from $13$ to $30$. For $W=6$, we find that upon increasing $N$
that there is a clear trend towards the GOE prediction. On the other
hand, the data for $W=12$ tend towards the Poissonian result. At
$W=9$, the $\Delta_3$ statistics is independent of $N$ within the
accuracy of our calculation. Thus the critical disorder for the
present example is $W_c\approx 9$. 
In addition to this finite-size dependence, we have also used finite
size scaling analysis for quantities such as
$\alpha=\int_0^{30}\Delta_3(L) dL$. We have constructed scaling
functions, which further support the values of $W_c$ obtained above
and confirm the one-parameter scaling hypothesis and thus the
existence of the MIT in these anisotropic systems.  Details will be
published elsewhere \cite{prepre}.

The critical disorders obtained by this analysis agree reasonably well
with the results from TMM \cite{Zamb96} and MFA \cite{Multi}. $W_c$
decreases with increasing anisotropy with a power law $W_c=16.3
(1-\gamma)^\beta$, where $\beta=0.25$ for weakly coupled planes and
$\beta\approx 0.6$ for weakly coupled chains, respectively.

For a given anisotropy and system size, we can also identify a
disorder $W'$ at which $P(s)$ agrees with the so called ``critical
statistics'', characteristic for the MIT of the isotropic system (cp.\ 
Fig.~\ref{fig:p_s}). However, this disorder is much smaller than $W_c$
for strong anisotropies. And, of course, the value of $W'$ changes
when we change the system size. Thus, $P_c(s)$ is not characteristic
for the MIT in anisotropic systems. We find that upon increasing the
anisotropy that the statistical properties at the MIT drift slowly
from the ``critical statistics'' $P_c(s)$ of the isotropic case
\cite{Zara97} towards Poisson statistics. In that sense, the states at
$W_c$ seem to be less extended in the anisotropic system compared to
the isotropic case.  This coincides with the results of the MFA
\cite{Multi}.
\begin{figure}[h]
  {\epsfysize=12cm\epsfbox{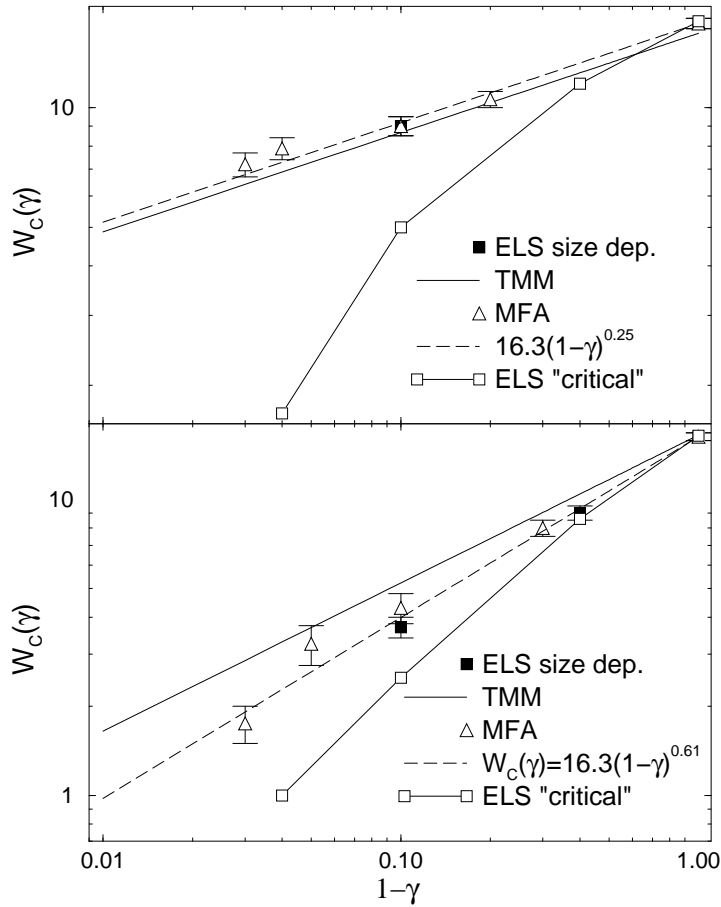}}
  \caption{\protect\small 
    $W_c(\gamma)$ for weakly coupled planes (above) and chains (below)
    as obtained with various methods. The thick solid line is the TMM
    result of Ref.~\protect\cite{Zamb96}. The open squares are the
    disorder values, where the ''critical statistics'' is found for
    system size $N=21$.}
\label{fig:wcPlanes}
\end{figure}

\section{Conclusions}
We find that a metal to insulator transition exists in the anisotropic
Anderson model. The critical disorders obtained from the ELS coincides
reasonably well with the results from TMM \cite{Zamb96} and MFA
\cite{Multi}. The system-size independent $P(s)$ and $\Delta_3(L)$ at
the MIT depend on the specific values chosen for the microscopic
hopping elements $t_x$, $t_y$, and $t_z$. They are different for each
$\gamma$ and the two anisotropy realizations, namely, weakly-coupled
planes and chains.  Furthermore, the ELS at the MIT is also different
from the ELS of the isotropic case. Thus we find that $P(s)$ at the
MIT is not universal,i.e., not independent of the microscopic
parameters of the model.
    \vspace{0.6cm}\newline{\small 
      We gratefully acknowledge support by the DFG within the
      Sonderforschungsbereich 393.
    }
    \end{document}